\documentclass[11pt,twoside]{article}
\usepackage{asp2014}

\aspSuppressVolSlug
\resetcounters

\begin{document}

\title{Using script generators for pipeline prototyping}

% full name: Dirk Petry
\author{Dirk~Petry$^1$}
\affil{$^1$European Southern Observatory, Garching bei M\"{u}nchen, Germany; \email{dpetry@eso.org}}
% remove/add as you need

% remove/add authors as you need
\paperauthor{Dirk~Petry}{dpetry@eso.org}{0000-0002-8704-7690}{European Southern Observatory}{ALMA Regional Centre}{Garching}{}{85748}{Germany}

% leave these next few aindex lines commented for the editors to enable them. Use Aindex.py to generate them for yourself.
% first presenting author should be the first entry for bold-facing the author index page-reference

% remove/add as you need

% leave the ssindex lines commented for the editors to enable them, use Index.py to suggest yours
%\ssindex{data!pipelines}
%\ssindex{software!scripting}

% leave the ooindex lines commented for the editors to enable them, use ascl.py to suggest yours
%\ooindex{CASA, ascl:1107.013}
  
\begin{abstract}

Fully automated astronomical data calibration and imaging pipelines are difficult to develop 
without a good prototyping method which permits to bridge the time between observatory commissioning 
and the moment when the special features and possible problems of the data and their processing 
are fully understood.
In this paper I present a method which has worked well for the ALMA observatory and which is sufficiently
general to be transferable to most other projects. In short, the idea is to use a three-level data analysis 
software design (scriptable toolkit, script generator, automated pipeline)
and a corresponding timing of the software development which is ramping up the effort in three stages 
starting at the beginning of construction, at the beginning of commissioning, and at the 
end of commissioning respectively.
The important design pattern which I would like to underline here is the use of script generators
as prototypes for the automated pipeline.
  
\end{abstract}

\section{Introduction}

With the ever larger data output volume of modern astronomical observatories,
automated data processing pipelines are an indispensable part of any such project.
New observatory projects are now aware that data analysis (DA) software
needs a significant lead time and has to be budgeted for carefully.
But even with adequate financial means, it is nearly impossible for the DA software 
to be ready {\it and verified} the moment the observatory records its first ``real'' data.
Any complex and novel observatory/detector system will have its unexpected features
and quirks which cannot be foreseen and simulated ahead of time.
As these unexpected features of the real data are gradually becoming apparent and understood, the
DA software needs to be adapted and equipped with corresponding heuristics. This is the usual commissioning
phase which is to be kept as short as possible such that scientific results can be produced soon.
Usually, a compromise is made and the new observatory/instrument is made available to
scientific users with a reduced set of features which is then gradually expanded
as commissioning of additional features is completed.
But even for relatively small sets of new capabilities, the commissioning to the point
of the release of a fully automated pipeline (PL), can be a long process of many months
if not years because the work requires know-how which is only available to a small expert team
that cannot arbitrarily be enlarged.
In this paper, I am presenting an approach to prototyping the automated PL
which has helped the ALMA observatory \citep{Cortes2020} since 2012 to shorten the time-to-release
of new capabilities by ca. one year. Its usefulness is not confined to the realm of radio astronomy. 

\section{Three-stage development start-up}
At the beginning of the construction of a new observatory, the arrival of the first real data is years away.
At this point, the data model for the raw data needs to be defined and the creation of simulated raw data prepared.
Then the development of the DA software can proceed in the following three stages:

{\bf 1 - Scriptable toolkit:}
The first layer of functionality which needs to be provided for commissioning scientists and 
later science users of the observatory, is a kit of DA tools from which an arbitrary
sequence of calibration and analysis steps can be constructed. These tools need to be bound
to a convenient scripting language, e.g. Python. It needs to include efficient methods to
import raw data into variables/arrays of the scripting language, perform common operations
on the imported data, and derive calibration information which can subsequently be stored
and applied to the raw data producing calibrated data in a convenient and efficient storage format.
Furthermore tools for data visualization need to be provided including interactive GUIs
for exploring the raw and calibrated data.
Finally, tools for transforming the calibrated data into astronomical images can already
be developed in this first phase based purely on simulated data.

{\bf 2 - Script generator:}
Once the beginning of instrument commissioning comes in sight and the scriptable toolkit
has reached some maturity, the development of the second layer of functionality can commence:
a tool to create standard data processing scripts for the given instrument - the {\it script generator} (SG).

This tool can be understood as a prototype PL: it examines a given set of raw data and then
outputs a (nearly) complete processing script tailored to the given dataset and 
using the commands from the scriptable toolkit. The SG already contains many of the heuristics
which will later be at work in the automated PL but it (at least initially, in its first releases) 
doesn't attempt to do the full job. Sections of the process which are not yet fully understood or
which are too complex to automate, may still be omitted in the output script draft such that
users can complete it manually.

\articlefigure[width=.5\textheight]{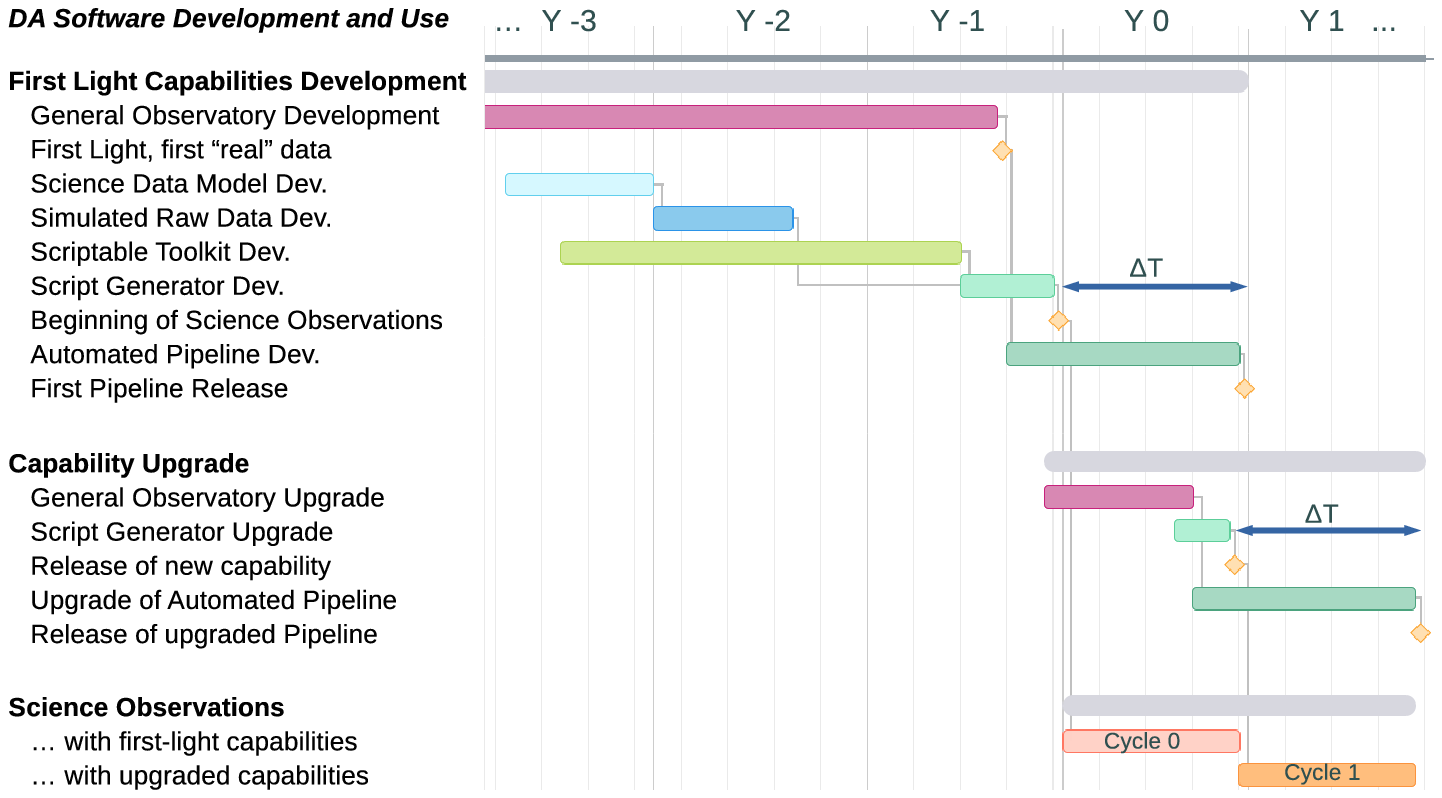}{processgant}{Simplified Gantt chart of the data analysis software development for an 
arbitrary observatory following the approach described in this paper. $\Delta T$ is the time period by which new capabilities 
can be released earlier because an SG was employed as a PL prototype. The exact time scale will depend on the project complexity. Continuation of the development of data model and toolkit is not shown.}

The resulting SG tool is a help for commissioning scientists and data analysts. 
It automates those parts of the data analysis process which are
already common knowledge, saving its users time that they can spend on the less well understood parts
of the analysis.

The development of an SG is not trivial but much simpler than that of a fully fledged PL.
Since the SG needs to read the instrument data, the SG needs to use the scriptable toolkit and 
therefore should be coded in the scripting language which is bound to it.
A first version of the SG can already be produced before the instrument's first light.
As soon as real data is available, the SG is rapidly updated with new heuristics.
Unit tests of the SG can be constructed from test datasets which exhibit certain instrument
features and capabilities. 

The execution time of an SG is typically much shorter than that of the subsequent execution of the DA script
which it produces because the SG mostly examines the {\it metadata} of the dataset, not the bulk data.
The automation of the scripting permits to produce fancy, standardized, well formated and marked-up 
scripts for the end-users.

At the end of the initial commissioning phase, when the observatory starts to take its first science data, 
the SG, like an expert system for writing data analysis scripts, encapsulates all the data processing knowledge 
which was accumulated by the commissioning scientists up to this point.
The SG can then be used for standard data processing of the observatory's science data
while the development team can give full attention to the third stage of development.

{\bf 3 - Automated pipeline:}
The final stage of the first round of DA software development is
the complete automation of the data calibration and science product generation 
(depending on the nature of the instrument this could, e.g., be imaging, the production of spectra etc.).
The fact that the SG is in place for the time being to provide a reasonably fast means
of processing data semi-automatically with the help of a number of data analysts,
buys time for the PL development team. 

This is the most labour-intensive phase of the project w.r.t. data processing. Three teams work in parallel:
(1) The commissioning scientists continue to monitor the most recent raw data and find further improvements
to the processing heuristics quickly updating them in the SG.
(2) Data analysts use the SG to generate data processing scripts which they run to produce the calibrated data
  and derived data products for the observatory archive and its users.
(3) The PL development team works on automating the process in order to speed it up and
  reduce the number of necessary data analysts in preparation for the ramping up of the output data volume of the observatory.
  Thorough testing on a diverse selection of datasets is needed.

Once the first official data processing PL is released, the work of the data analysts begins to change
from mostly working with the SG to more and more double-checking the PL products.
As confidence in the PL grows, the fraction of the total data volume which is handled by it
approaches 100\%. Now the observatory can afford to further increase its output data volume.
The SG is only used in exceptional cases, {\it and} during commissioning of new capabilities.

\section{Gradual addition of further capabilities}
For ground-based observatories, the set of capabilities can be extended indefinitely. Every time a new capability is added,
the commissioning scientists can fall back to semi-automatic data processing with the help of the SG. 
They develop new heuristics for the new type of data, and add them to the features of the SG.
Once the new capability is released to the observatory users, the observatory data analysts will at first use
the SG again to process this subset of the observatory output.
At the same time, the PL development team can start to enable the PL to handle the new type of data.
After one development cycle, the PL has learned to handle the new capability.

In this way, observatory users can be given access to well calibrated data and good derived data products from 
recently added observatory upgrades with a minimum delay.

\section{The process in real life: ALMA}
The data analysis software development approach described above is the path which the ALMA observatory
followed on its way into science operations and through many capability expansions.
The raw data model and format chosen by ALMA 
is the ALMA Science Data Model (ASDM, \citet{2006ASPC..351..627V}).
ALMA's scriptable data analysis toolkit is CASA \citep{2007ASPC..376..127M, 2020ASPC..527..267E} which 
uses the scripting language Python. The ALMA raw data was calibrated during the first years of operations starting 2012  
using exclusively the Calibration Script Generator (CSG)
 \citep[see][]{2014SPIE.9152E..0JP}, a tool also written in Python. Then, in 2014, the first ALMA calibration PL was released
(the documentation of this and subsequent PL versions can be found at
\url{https://almascience.org/processing/science-pipeline}, the latest is \citet{ALMAPipe2021}).
Since then, the CSG and the calibration PL have co-existed. ALMA data is calibrated
by the calibration PL whenever possible. For new observing modes, the CSG is upgraded first and used for ca. 
one year of operations while the PL is being upgraded correspondingly. After that, the PL takes over the processing of the new mode.
Since ALMA also aims to offer high-quality images to its users, there is a second processing step after calibration to
produce these. This was initially done using imaging script templates. Then a second script generator was created,
the Imaging Script Generator (ISG). In 2016, the first ALMA PL with automated imaging was released. 
Today, 95\% of the ALMA data are PL calibrated and imaged \citep{2020SPIE11449E..1TN}. 
CSG and ISG have been released in 2021 for public use 
(see \url{https://confluence.alma.cl/display/EAPR/ALMA+Data+Analysis+Utilities}).

\bibliography{X9-005}

\begin{thebibliography}{}
\expandafter\ifx\csname natexlab\endcsname\relax\def\natexlab#1{#1}\fi
\expandafter\ifx\csname url\endcsname\relax
  \def\url#1{\texttt{#1}}\fi
\expandafter\ifx\csname urlprefix\endcsname\relax\def\urlprefix{URL }\fi
\providecommand{\eprint}[2][]{\url{#2}}

\bibitem[{{ALMA Pipeline Team}(2021)}]{ALMAPipe2021}
{ALMA Pipeline Team} 2021, {ALMA Science Pipeline User's Guide, ALMA Doc
  2021.2v1.0}

\bibitem[{{Cortes} et~al.(2020){Cortes}, {Remijan}, {Biggs}, {Dent},
  {Carpenter}, {Fomalont}, {Hales}, {Kameno}, {Mason}, {Philips}, {Saini},
  {Stoehr}, {Vila-Vilaro}, \& {Villard}}]{Cortes2020}
{Cortes}, P.~C., {Remijan}, A., {Biggs}, A., {Dent}, B., {Carpenter}, J.,
  {Fomalont}, E., {Hales}, A., {Kameno}, S., {Mason}, B., {Philips}, N.,
  {Saini}, K., {Stoehr}, F., {Vila-Vilaro}, B., \& {Villard}, E. 2020, {ALMA
  Technical Handbook, ALMA Doc. 8.5, ver. 1.0}

\bibitem[{{Emonts} et~al.(2020){Emonts}, {Raba}, {Moellenbrock}
  et~al.}]{2020ASPC..527..267E}
{Emonts}, B., {Raba}, R., {Moellenbrock}, G., et~al. 2020, in ADASS XXIX,
  edited by R.~{Pizzo}, et~al., vol. 527 of ASP Conf. Ser., 267

\bibitem[{{McMullin} et~al.(2007){McMullin}, {Waters}, {Schiebel}, {Young}, \&
  K.}]{2007ASPC..376..127M}
{McMullin}, J., {Waters}, B., {Schiebel}, D., {Young}, W., \& K., G. 2007, in
  ADASS XVI, edited by F.~{Shaw}, R.A.~{Hill}, \& D.~{Bell}, vol. 376 of ASP
  Conf. Ser., 127

\bibitem[{{Nakos} et~al.(2020){Nakos}, {Francke}, {Nakanishi}, {Petry},
  {Stanke}, {Ubach}, {Cerrigone}, {Keller}, {Trejo}, \&
  {Ueda}}]{2020SPIE11449E..1TN}
{Nakos}, T., {Francke}, H., {Nakanishi}, K., {Petry}, D., {Stanke}, T.,
  {Ubach}, C., {Cerrigone}, L., {Keller}, E., {Trejo}, A., \& {Ueda}, J. 2020,
  in Observatory Operations: Strategies, Processes, and Systems VIII, vol.
  11449 of SPIE Astronomical Telescopes and Instrumentation

\bibitem[{{Petry} et~al.(2014){Petry}, {Vila-Vilaro}, {Villard}, {Komugi}, \&
  {Schnee}}]{2014SPIE.9152E..0JP}
{Petry}, D., {Vila-Vilaro}, B., {Villard}, E., {Komugi}, S., \& {Schnee}, S.
  2014, in Software and Cyberinfrastructure for Astronomy III, edited by
  G.~{Chiozzi}, \& N.~{Radziwill}, vol. 9152 of SPIE Astronomical Telescopes
  and Instrumentation

\bibitem[{{Viallefond}(2006)}]{2006ASPC..351..627V}
{Viallefond}, F. 2006, in ADASS XV, edited by C.~{Gabriel}, C.~{Arviset},
  D.~{Ponz}, \& E.~{Solano}, vol. 351 of ASP Conf. Ser., 627

\end{thebibliography}

\end{document}